\documentclass[12pt]{article}
\usepackage{graphicx}
\usepackage{latexsym}
\usepackage{amscd}
\usepackage[cp1251]{inputenc}
\usepackage[english,russian]{babel}
\usepackage{amsmath,  eucal}
\usepackage{amssymb}
\usepackage{indentfirst}
\usepackage[small,centerlast]{caption2}
\usepackage{longtable}
\usepackage[dvipsnames]{color}% - это необходимо для правки
\usepackage{cite}
\graphicspath{{figure/}}
\makeatletter
\newcommand{\const}{\mathop{\rm const\, }}

\makeatother {

\makeatletter
\renewcommand{\section}{\@startsection {section}{1}{\z@}%
                                   {-3.5ex \@plus -1ex \@minus -.2ex}%
                                   {2.3ex \@plus.2ex}%
                                   {\normalfont\Large\uppercase}}
\renewcommand{\subsection}{\@startsection{subsection}{2}{\z@}%
                                     {-3.25ex\@plus -1ex \@minus -.2ex}%
                                     {1.5ex \@plus .2ex}%
                                     {\normalfont\large\itshape}}
\renewcommand{\subsubsection}{\@startsection{subsubsection}{3}{1em}%
                                     {-3.25ex\@plus -1ex \@minus -.2ex}%
                                     {-1.5em \@plus .2em}%
                                     {\normalfont\normalsize\bfseries}}

\makeatother

\begin{document}
\renewcommand{\refname}{\begin{center}\bf REFERENCES\end{center}}
\newcommand{\mc}[1]{\mathcal{#1}}
\newcommand{\E}{\mc{E}}
\thispagestyle{empty} \large
\renewcommand{\abstractname}{ Abstract}

 \begin{center}
{\textbf{Longitudinal electric current in classical
collisional Maxwellian plasma, inducted by transversal electromagnetic wave}}
\medskip
\end{center}

\begin{center}
  \bf A. V. Latyshev\footnote{$avlatyshev@mail.ru$} and
  A. A. Yushkanov\footnote{$yushkanov@inbox.ru$}
\end{center}\medskip

\begin{abstract}
Kinetic Vlasov equation for collisional Maxwellian plasmas is used.
Collision  integral of BGK (Bhatnagar, Gross and Krook) type is applied.
From Vlasov equation we find  distribution  function of electrons
in square-law approximation on size of transversal electric field.
The formula for electric current calculation is deduced.
This formula contains an one-dimensional quadrature.
It is shown, that nonlinearity leads to revealing of the longitudinal
electric current directed along a wave vector.
This longitudinal current is perpendicular to known so-called
transversal classical current.
The classical current turns out at the linear analysis.
The longitudinal current  in case of small values of wave numbers
is calculated.
When frequency of collisions tends to zero, all received
formulas for collisional  plasmas pass in the known
corresponding results for collisionless plasmas.
Graphic research of dimensionless density of a current is carried
out.
\end{abstract}

\section*{\bf Introduction}

В настоящей работе выводятся формулы для вычисления
электрического тока в максвелловской столкновительной плазме.
В качестве интеграла столкновений используется
известный интеграл столкновений БГК (Бхатнагар, Гросс и Крук)
(см., \cite{BGK} и \cite{Welander}).

При решении кинетического уравнения Власова, описывающего
поведение плазмы, мы учитываем величины, пропорциональные квадрату
напряженности внешнего электрического поля.
При этом мы используем квадратичные разложения
функции распределения, самосогласованного электромагнитного поля
и интеграла столкновений.

При таком нелинейном подходе оказалось, что электрический ток
имеет две ненулевые компоненты. Одна компонента электрического
тока направлена вдоль напряженности электрического поля и пропорциональна
его величине.
Эта компонента электрического тока в точности та же самая, что и
в линейном анализе. Это "поперечный"\, ток. Следовательно, в
линейном анализе мы получаем известное выражение поперечного
электрического тока.

Вторая ненулевая компонента электрического тока имеет второй
порядок малости относительно напряженности электрического тока
и пропорциональна его квадрату.
Вторая компонента электрического тока направлена вдоль волнового
вектора перпендикулярно напряженности электрического поля. Этот
ток ортогонален первой компоненте тока. Это "продольный"\, ток.

Генерирование в плазме продольного тока поперечным электромагнитным
полем выявляется  нелинейным анализом
взаимодействия электромагнитного поля с плазмой.

Нелинейные эффекты в плазме изучаются длительное время
\cite{Gins} -- \cite{Shukla1}.

В работе \cite{Zyt} нелинейный ток был использован, в частности,
в вопросах вероятности распадных процессов.
Заметим, что в работе \cite{Zyt2} отмечено существование
нелинейного тока вдоль волнового вектора
(см. формула  (2.9) из \cite{Zyt2}).

Квантовая плазма интенсивно изучалась в последние десятилетия
(см., например, \cite{Lat1} - \cite{Lat9}).
Столкновительная квантовая плазма начала изучаться в работе
\cite{Mermin}. Затем квантовая столкновительная плазма
изучалась в наших работах \cite{Lat2} - \cite{Lat6}.
В этих работах изучалась квантовая столкновительная плазма
с произвольной переменной частотой столкновений.

В работах \cite{Lat7} - \cite{Lat9} было исследовано
генерирование продольного тока поперечным электромагнитным полем
электромагнитным полем в классической и квантовой плазме
Ферми---Дирака \cite{Lat7}, в максвелловской плазме
\cite{Lat8} и в вырожденной плазме \cite{Lat9}.

В настоящей работе выведены формулы для вычисления продольного
тока, генерируемого поперечным электромагнитным полем, в
максвелловской столкновительной плазме.

\section{\bf Решение уравнения Власова}

Покажем, что в максвелловской плазме, описываемой уравнением
Власова, генерируется продольный ток поперечным электрическим
полем и вычислим его плотность.
На существование этого тока указывалось более полувека тому
назад \cite{Zyt2}.

Возьмем уравнение Власова, описывающее поведение
столкновительной плазмы с интегралом столкновений БГК (Бхатнагар, Гросс и
Крук) \medskip
$$
\dfrac{\partial f}{\partial t}+\mathbf{v}\dfrac{\partial f}{\partial
\mathbf{r}}+
e\bigg(\mathbf{E}+
\dfrac{1}{c}[\mathbf{v},\mathbf{H}]\bigg)
\dfrac{\partial f}{\partial\mathbf{p}}=\nu(f^{(0)}-f).
\eqno{(1.1)}
$$ \medskip

В уравнении (1.1) $f$ -- функция распределения электронов плазмы,
${\bf E}, {\bf H}$ -- компоненты электромагнитного поля, $c$
-- скорость света, ${\bf p}=m{\bf v}$ импульс электронов,
${\bf v}$ -- скорость электронов, $\nu$ -- эффективная частота столкновений
электронов с  частицами плазмы,
$f^{(0)}=f_{eq}({\bf r},v)$ (eq $\equiv$ equilibrium)
-- локально равновесная функция распределения Максвелла---Больцмана,\medskip
$$
f_{eq}({\bf r},v)=v_T^{-3}\pi^{-3/2}N({\bf r},t)\exp\Big(-\dfrac{\E}{k_BT}\Big),
$$
\medskip
$\E={mv^2}/{2}$ -- энергия электронов,  $k_B$ -- постоянная Больцмана, $T$
-- температура плазмы,
${\bf P}=\dfrac{{\bf P}}{p_T}=\dfrac{{\bf v}}{v_T}$ -- безразмерный импульс
(скорость) электронов,  $p_T=mv_T$,
$v_T$ -- тепловая скорость электронов,
$$
v_T=\sqrt{\dfrac{2k_BT}{m}},\qquad k_BT=\E_T=\dfrac{mv_T^2}{2}
$$
-- тепловая энергия электронов.

Равновесная функция распределения обладает свойством (нормировка
на числовую плотность)
$$
\int f_{eq}({\bf r},v)d^3v=N({\bf r},t).
$$

Будем считать, что в плазме имеется электромагнитное поле,
представляющее собой бегущую гармоническую волну
$$
{\bf E}={\bf E}_0e^{i({\bf kr}-\omega t)}, \qquad
{\bf H}={\bf H}_0e^{i({\bf kr}-\omega t)}.
$$

Электрическое и магнитное поля связаны между собой векторным
потенциалом следующими равенствами
$$
\mathbf{E}=-\dfrac{1}{c}\dfrac{\partial \mathbf{A}}{\partial t},
\;\qquad
\mathbf{H}={\rm rot} \mathbf{A}.
$$

Для определенности будем считать, что волновой вектор направлен
вдоль оси $x $, в электрическое поле направлено вдоль оси $y $,
т.е.
$$
{\bf k}=k(1,0,0), \qquad {\bf E}=E_y(x,t)(0,1,0).
$$

Следовательно,
$$
\mathbf{E}=-\dfrac{1}{c}\dfrac{\partial \mathbf{A}}{\partial t}
=\dfrac{i\omega}{c}\mathbf{A},
$$ \medskip
$$
{\bf H}=\dfrac{ck}{\omega}E_y\cdot(0,0,1),\qquad
{\bf [v,H}]=\dfrac{ck}{\omega}E_y\cdot (v_y,-v_x,0),
$$ \medskip
$$
e\bigg(\mathbf{E}+\dfrac{1}{c}[\mathbf{v},\mathbf{H}]\bigg)
\dfrac{\partial f}{\partial\mathbf{p}}=
\dfrac{e}{\omega}E_y\Big[kv_y\dfrac{\partial f}{\partial p_x}+
(\omega-kv_x)\dfrac{\partial f}{\partial p_y}\Big],
$$ \medskip
а также
$$
[\mathbf{v,H}]\dfrac{\partial f_0}{\partial \mathbf{p}}=0,\quad
\text{так как}\quad
\dfrac{\partial f_0}{\partial \mathbf{p}}\sim \mathbf{v}.
$$

Рассмотрим линеаризацию локально равновесной функции
распределения
$$
f_{eq}(P,x)=f_0(P)+f_0(P)\dfrac{\delta N}{N},
$$
где
$$
f_0(P)=v_T^{-3}\pi^{-3/2}N\exp\Big(-\dfrac{\E}{k_BT}\Big)=
\dfrac{N}{v_T^3\pi^{3/2}}e^{-P^2},
$$
$$
N(x,t)=N+\delta N(x,t),\qquad N=\const.
$$

Уравнение (1.1) может быть переписано в виде
$$
\dfrac{\partial f}{\partial t}+v_x\dfrac{\partial f}{\partial x}+
\dfrac{eE_y}{\omega}\Big[kv_y\dfrac{\partial f}{\partial p_x}+
(\omega-kv_x)\dfrac{\partial f}{\partial p_y}\Big]+\nu f=
$$
$$
=\nu f_0(P)+\nu \dfrac{\delta N}{N}f_0(P).
\eqno{(1.2)}
$$

Величина $ \delta N/N $ может быть найдена из закона
сохранения числа частиц
$$
\int (f_{eq}-f)d^3v=0.
$$

Из этого закона сохранения мы получаем
$$
\dfrac{\delta N}{N}\int f_0(P)d^3v=\int
(f-f_0(P))d^3v.
$$
Из этого уравнения получаем, что
$$
\delta N=\int (f-f_0(P))d^3v=v_T^3\int(f-f_0(P))d^3P.
$$

Уравнение (1.2) может быть преобразовано теперь к интегральному уравнению
$$
\dfrac{\partial f}{\partial t}+v_x\dfrac{\partial f}{\partial x}+
\nu f=\nu f_0(P)-\dfrac{eE_y}{\omega}\Big[kv_y\dfrac{\partial f}{\partial p_x}+
(\omega-kv_x)\dfrac{\partial f}{\partial p_y}\Big]+
$$
$$
+\nu f_0(P)\dfrac{v_T^3}{N}\int [f-f_0(P)] d^3P.
\eqno{(1.3)}
$$

Будем искать решение уравнения (1.3) в виде
$$
f=f_0(P)+f_1+f_2,
\eqno{(1.4)}
$$
где
$$
f_1\sim E_y\sim e^{i(kx-\omega t)},
$$
$$
f_2\sim E_y^2\sim e^{2i(kx-\omega t)}.
$$

Будем действовать методом последовательных приближений,
считая малым параметром величину напряженности электрического
поля.
Уравнение (1.3) с помощью (1.4) теперь можно переписать в виде:
$$
\dfrac{\partial (f_1+f_2)}{\partial t}+v_x\dfrac{\partial (f_1+f_2)}{\partial x}+
\nu (f_1+f_2)=$$$$=-\dfrac{eE_y}{\omega}\Big[kv_y\dfrac{\partial (f_0(P)+f_1)}
{\partial p_x}+(\omega-kv_x)\dfrac{\partial (f_0(P)+f_1)}{\partial p_y}\Big]+
$$
$$
+\nu f_0(P)\dfrac{v_T^3}{N}\int [f_1+f_2] d^3P.
\eqno{(1.3')}
$$

Теперь уравнение (1.3)  эквивалентно следующим двум уравнениям
$$
\dfrac{\partial f_1}{\partial t}+
v_x\dfrac{\partial f_1}{\partial x}+\nu f_1=
$$
$$
=-\dfrac{eE_y}{\omega}\Bigg[kv_y\dfrac{\partial f_0(P)}{\partial p_x}+
(\omega-kv_x)\dfrac{\partial f_0(P)}{\partial p_y}\Bigg]
+\nu f_0(P)\dfrac{v_T^3}{N}\int f_1 d^3P.
\eqno{(1.5)}
$$ \bigskip
и
$$
\dfrac{\partial f_2}{\partial t}+
v_x\dfrac{\partial f_2}{\partial x}+\nu f_2=
$$
$$
=-\dfrac{eE_y}{\omega}\Bigg[kv_y\dfrac{\partial f_1}{\partial p_x}+
(\omega-kv_x)\dfrac{\partial f_1}{\partial p_y}\Bigg]
+\nu f_0(P)\dfrac{v_T^3}{N}\int f_2 d^3P.
\eqno{(1.6)}
$$ \bigskip

Из уравнения (1.5) получаем
$$
(\nu-i\omega+ikv_x)f_1=
$$
$$
=-\dfrac{eE_y}{\omega}
\Bigg[kv_y\dfrac{\partial f_0(P)}{\partial p_x}+
(\omega-kv_x)\dfrac{\partial f_0(P)}{\partial p_y}\Bigg]
+\nu f_0(P)\dfrac{v_T^3}{N} A_1.
$$

Здесь
$$
A_1=\int f_1 d^3P.
\eqno{(1.7)}
$$

Введем безразмерные параметры
$$
\Omega=\dfrac{\omega}{k_Tv_T},\qquad y=\dfrac{\nu}{k_Tv_T},
\qquad q=\dfrac{k}{k_T}.
$$

Здесь $q $ безразмерное волновое число,
$k_T =\dfrac {mv_T} {\hbar} $ -- тепловое волновое число, $ \Omega $
-- безразмерная частота колебаний электромагнитного поля.

В предыдущем уравнении перейдем к безразмерным параметрам
$$
i(qP_x-z)f_1=$$$$=-\dfrac{eE_y}{\Omega k_Tp_Tv_T}
\Bigg[qP_y\dfrac{\partial f_0(P)}{\partial P_x}+
(\Omega-qP_x)\dfrac{\partial f_0(P)}{\partial P_y}\Bigg]+
y f_0(P)\dfrac{v_T^3}{N}A_1.
\eqno{(1.8)}
$$

Здесь
$$
z=\Omega+iy=\dfrac{\omega+iy}{k_Tv_T}.
$$

Заметим, что
$$
\dfrac{\partial f_0(P)}{\partial P_x}\sim P_x,\qquad
\dfrac{\partial f_0(P)}{\partial P_y}\sim P_y.
$$

Следовательно
$$
\Bigg[qP_y\dfrac{\partial f_0(P)}{\partial P_x}+
(\Omega-qP_x)\dfrac{\partial f_0(P)}{\partial P_y}\Bigg]=
\Omega\dfrac{\partial f_0(P)}{\partial P_y}.
$$

Теперь из уравнения (1.8) находим, что
$$
f_1=\dfrac{ieE_y}{k_Tp_Tv_T}\cdot\dfrac{\partial f_0/\partial P_y}
{qP_x-z}-iy\dfrac{v_T^3}{N}\cdot\dfrac{f_0(P)}{qP_x-z}A_1.
\eqno{(1.9)}
$$

Подставляя (1.9) в уравнение (1.7), получаем равенство
$$
A_1\Bigg(1+iy \dfrac{v_T^3}{N}\int \dfrac{f_0(P)d^3P}
{qP_x-z}\Bigg)=\dfrac{ieE_y}{k_Tp_Tv_T}\int
\dfrac{\partial f_0/\partial P_y}{qP_x-z}d^3P.
$$

Легко видеть, что интеграл в правой части этого равенства равен
нулю. Следовательно, $A_1=0$. Следовательно, согласно (1.9)
функция $f_1$  построена и определяется равенством
$$
f_1=\dfrac{ieE_y}{k_Tp_Tv_T}\cdot\dfrac{\partial f_0/\partial P_y}
{qP_x-z}.
\eqno{(1.10)}
$$

Во втором приближении подставим $f_1 $ согласно (1.10)
в уравнение (1.6).

Получим уравнение
$$
(\nu-2i\omega+2ikv_x)f_2=$$$$
-\dfrac{ie^2E_y^2}{k_Tp_Tv_T\omega}\Big[kv_y\dfrac{\partial}{\partial p_x}
\Big(\dfrac{\partial f_0/\partial P_y}{qP_x-z}\Big)+
(\omega-kv_x)\dfrac{\partial }
{\partial p_y}\Big(\dfrac{\partial f_0/\partial
P_y}{qP_x-z}\Big)\Big]+
$$
$$
+\nu f_0(P)\dfrac{v_T^3}{N}A_2.
$$

Здесь
$$
A_2=\int f_2d^3P.
\eqno{(1.11)}
$$

Перейдем в этом уравнении к безразмерным параметрам. Получим
уравнение
$$
2i(qP_x-x-\dfrac{iy}{2})f_2=$$$$=-\dfrac{ie^2E_y^2}{\Omega k_T^2p_T^2v_T^2}
\Big[qP_x\dfrac{\partial}{\partial P_x}
\Big(\dfrac{\partial f_0/\partial P_y}{qP_x-z}\Big)+
(\Omega-qP_x)\dfrac{\partial }
{\partial P_y}\Big(\dfrac{\partial f_0/\partial
P_y}{qP_x-z}\Big)\Big]+
$$
$$
+y f_0(P)\dfrac{v_T^3}{N}A_2.
$$
Обозначим
$$
z'=\Omega+\dfrac{iy}{2}=\dfrac{\omega}{k_Tv_T}+i\dfrac{\nu}{2k_Tv_T}=
\dfrac{\omega+i \nu/2}{k_Tv_T}.
$$

Из последнего уравнения находим
$$
f_2=-\dfrac{e^2E_y^2}{2k_T^2p_T^2v_T^2 \Omega}
\Bigg[qP_y\dfrac{\partial}{\partial P_x}
\Big(\dfrac{\partial f_0/\partial P_y}{qP_x-z}\Big)+
\dfrac{\Omega-qP_x}{qP_x-z}\dfrac{\partial^2f_0}{\partial P_y^2}\Bigg]
\dfrac{1}{qP_x-z'}-
$$
$$
-\dfrac{iy}{2}\cdot\dfrac{v_T^3}{N}\dfrac{f_0(P)}{qP_x-z'}A_2.
\eqno{(1.12)}
$$

Для нахождения $A_2$ подставим (1.12) в (1.11). Из полученного
соотношения находим $A_2$
$$
A_2=-\dfrac{e^2E_y^2}{2k_T^2p_T^2v_T^2\Omega}\cdot\dfrac{J_1}
{1+\dfrac{iy}{2}\dfrac{v_T^3}{N}J_0}.
$$

Здесь
$$
J_0=\int\dfrac{f_0(P)d^3P}{qP_x-z'},
$$
$$
J_1=\int\Bigg[qP_y\dfrac{\partial}{\partial P_x}
\Big(\dfrac{\partial f_0/\partial P_y}{qP_x-z}\Big)+
\dfrac{\Omega-qP_x}{qP_x-z}\dfrac{\partial^2f_0}{\partial P_y^2}\Bigg]
\dfrac{d^3P}{qP_x-z'}.
$$

Подставляя $A_2$ в (1.12), окончательно находим $f_2$ в явном виде
$$
f_2=-\dfrac{e^2E_y^2}{2k_T^2p_T^2v_T^2 \Omega}
\Bigg[qP_y\dfrac{\partial}{\partial P_x}
\Big(\dfrac{\partial f_0/\partial P_y}{qP_x-z}\Big)+
\dfrac{\Omega-qP_x}{qP_x-z}\dfrac{\partial^2f_0}{\partial P_y^2}\Bigg]
\dfrac{1}{qP_x-z'}+
$$
$$
+\gamma\dfrac{e^2E_y^2}{2k_T^2p_T^2v_T^2 \Omega}\cdot\dfrac{v_T^3}{N}
\cdot\dfrac{f_0(P)}{qP_x-z'}.
\eqno{(1.13)}
$$

Здесь
$$
\gamma=\dfrac{(iy/2)J_1}{1+(iy/2)(v_T^3/N)J_0}.
\eqno{(1.14)}
$$

\section{Плотность электрического тока}

Найдем плотность электрического тока
$$
\mathbf{j}=e\int \mathbf{v}f d^3v.
\eqno{(2.1)}
$$

Из равенств (1.4) -- (1.6) видно, что вектор плотности тока
имеет две компоненты
$$
\mathbf{j}=(j_x,j_y,0).
$$

Здесь $j_y$ -- плотность поперечного тока,
$$
j_y=e\int v_yf d^3v=e\int v_yf_1 d^3v,
$$
или
$$
j_y=e\int v_yf_1 d^3v=ev_T^4\int P_y f_1 d^3P.
$$

Этот ток направлен вдоль электрического поля, его плотность
определяется только первым приближением функции распределения.

Второе приближение функции распределеня не вносит вклад в
плотность тока.

Плотность поперечного тока определяется равенством
$$
j_y=\dfrac{ie^2v_T^3}{k_Tp_T}E_y(x,t)
\int\dfrac{(\partial f_0/\partial P_y)P_y}{qP_x-z}d^3P.
$$

Этот ток пропорционален первой степени величины напряженности
электрического поля.

Для плотности продольного тока по его определению имеем
$$
j_x=e\int v_xf d^3v=
e\int v_xf_2 d^3v=ev_T^4\int P_xf_2d^3P.
$$

С помощью (1.13) отсюда получаем
$$
j_x=\dfrac{e^3E_y^2v_T^2}{2k_T^2p_T^2\Omega}\Bigg[-\int
\Bigg[qP_y\dfrac{\partial}{\partial P_x}
\Big(\dfrac{\partial f_0/\partial P_y}{qP_x-z}\Big)+
\dfrac{x-qP_x}{qP_x-z}\cdot\dfrac{\partial^2f_0}{\partial P_y^2}\Bigg]
\dfrac{P_xd^3P}{qP_x-z'}+
$$
$$
+\gamma \dfrac{v_T^3}{N}\int\dfrac{P_xf_0(P)d^3P}{qP_x-z'}
\Bigg].
\eqno{(2.2)}
$$

В интеграле от второго слагаемого из квадратных скобок (2.2)
внутренний интеграл по $P_y $ равен нулю
$$
\int\limits_{-\infty}^{\infty}\dfrac{\partial^2f_0}{\partial P_y^2}dP_y
=\dfrac{\partial f_0}{\partial P_y}\Bigg|_{P_y=-\infty}^{P_y=+\infty}=0.
$$

В первом интеграле из квадратных скобок (2.2) внутренний интеграл по $P_x $
вычисляется по частям
$$
\int\limits_{-\infty}^{\infty}\dfrac{\partial}{\partial P_x}
\Big(\dfrac{\partial f_0/\partial P_y}{qP_x-z}\Big)
\dfrac{P_xdP_x}{qP_x-z'}=z' \int\limits_{-\infty}^{\infty}
\dfrac{\partial f_0/\partial P_y}{(qP_x-z)(qP_x-z')^2}dP_x.
$$

Следовательно, равенство (2.2) упрощается
$$
j_x=\dfrac{e^3E_y^2v_T^2}{2k_T^2p_T^2\Omega}\Bigg[-z'q\int
\dfrac{P_y(\partial f_0/\partial P_y)d^3P}{(qP_x-z)(qP_x-z')^2}+
$$
$$
+\gamma \dfrac{v_T^3}{N}\int\dfrac{P_x f_0(P)d^3P}{qP_x-z'}\Bigg].
$$

Внутренний интеграл по переменной $P_y $ вычислим по частям
$$
\int\limits_{-\infty}^{\infty}P_y\dfrac{\partial f_0}{\partial P_y}dP_y=
P_yf_0\Bigg|_{P_y=-\infty}^{P_y=+\infty}-
\int\limits_{-\infty}^{\infty}f_0(P)dP_y=
-\int\limits_{-\infty}^{\infty}f_0(P)dP_y.
$$

Следовательно,  получим выражение для продольного тока \medskip
$$
j_x=\dfrac{e^3E_y^2v_T^2}{2k_T^2p_T^2\Omega}\Bigg[z'q\int
\dfrac{f_0(P)d^3P}{(qP_x-z)(qP_x-z')^2}+
\gamma \dfrac{v_T^3}{N}\int\dfrac{P_x f_0(P)d^3P}{qP_x-z'}\Bigg].
\eqno{(2.3)}
$$ \medskip

Внутренний интеграл в плоскости $ (P_y, P_z) $ равен:
$$
\int\dfrac{f_0(P)d^3P}{(qP_x-z)(qP_x-z')^2}=\int\limits_{-\infty}^{\infty}
\dfrac{dP_x}{(qP_x-z)(qP_x-z')^2}
\int\limits_{-\infty}^{\infty}\int\limits_{-\infty}^{\infty}
f_0(P)dP_ydP_z=
$$
$$
=\dfrac{N}{v_T^3 \sqrt{\pi}}\int\limits_{-\infty}^{\infty}
\dfrac{e^{-P_x^2}dP_x}{(qP_x-z)(qP_x-z')^2}=
\dfrac{N}{v_T^3}J_{12},
$$
где
$$
J_{12}=\dfrac{1}{\sqrt{\pi}}\int\limits_{-\infty}^{\infty}
\dfrac{e^{-P_x^2}dP_x}{(qP_x-z)(qP_x-z')^2}.
$$

Кроме того,
$$
\dfrac{v_T^3}{N}\int\dfrac{P_xf_0(P)d^3P}{qP_x-z'}=
\dfrac{v_T^3}{N}\pi\int\limits_{-\infty}^{\infty}
\dfrac{P_xf_0(P_x)dP_x}{qP_x-z'}=
\dfrac{1}{\sqrt{\pi}}
\int\limits_{-\infty}^{\infty}\dfrac{e^{-\tau^2}\tau d \tau}
{q\tau-z'},
$$

Обозначим
$$
J_{02}=\dfrac{1}{\sqrt{\pi}}\int\limits_{-\infty}^{\infty}
\dfrac{e^{-\tau^2}\tau d \tau}{q\tau-z'}.
$$

Равенство (2.3) сводится к следующему
$$
j_x=\dfrac{e^3E_y^2v_T^2}{2k_T^2p_T^2\Omega}
\Bigg[z'q \dfrac{N}{v_T^3}J_{12}+\gamma J_{02}\Bigg].
\eqno{(2.4)}
$$

Вернемся к рассмотрению величины $\gamma$. Вычислим интегралы,
входящие в (1.14). Вычислим первый интеграл:
$$
J_1=\int
\Bigg[qP_y\dfrac{\partial}{\partial P_x}
\Big(\dfrac{\partial f_0/\partial P_y}{qP_x-z}\Big)+
\dfrac{x-qP_x}{qP_x-z}\dfrac{\partial^2f_0}{\partial P_y^2}\Bigg]
\dfrac{d^3P}{qP_x-z'}.
$$
Как уже указывалось, интеграл от второго слагаемого равен нулю.
Второй интеграл как и ранее вычислим по частям. В результате
получаем:
$$
J_1=q\int P_y \dfrac{\partial}{\partial P_x}
\Big(\dfrac{\partial f_0/\partial P_y}{qP_x-z}\Big)\dfrac{d^3P}{qP_x-z'}=
$$
$$
=q^2\int \dfrac{P_y[\partial f_0/\partial P_y]d^3P}{(qP_x-z)(qP_x-z')^2}.
$$
Теперь вычислим по частям внутренний интеграл по переменной
$P_y$. В результате получаем:
$$
J_1=-q^2\int\dfrac{ f_0(P)d^3P}{(qP_x-z)(qP_x-z')^2}.
$$
Этот интеграл был вычислен ранее. Следовательно,
$$
J_1=-q^2\dfrac{N}{v_T^3\sqrt{\pi}}\int\limits_{-\infty}^{\infty}
\dfrac{e^{-\tau^2}d\tau}{(q\tau-z)(q\tau-z')^2}=
-q^2 \dfrac{N}{v_T^3} J_{12}.
$$

Вычислим второй интеграл из (1.14). Имеем:
$$
J_0=\int\dfrac{f_0(P)d^3P}{qP_x-z'}=
\dfrac{N}{v_T^3\sqrt{\pi}}\int\limits_{-\infty}^{\infty}
\dfrac{e^{-P_x^2}dP_x}{qP_x-z'}.
$$

Следовательно,
$$
1+\dfrac{iy}{2}\dfrac{v_T^3}{N}J_0=1+\dfrac{iy}{2}\dfrac{1}{\sqrt{\pi}}
\int\limits_{-\infty}^{\infty}\dfrac{e^{-\tau^2}d\tau}{q\tau-z'}=
$$
$$
=\dfrac{1}{\sqrt{\pi}}\int\limits_{-\infty}^{\infty}
e^{-\tau^2}d\tau+\dfrac{iy}{2}
\dfrac{1}{\sqrt{\pi}}
\int\limits_{-\infty}^{\infty}\dfrac{e^{-\tau^2}d\tau}{q\tau-z'}=
\dfrac{1}{\sqrt{\pi}}\int\limits_{-\infty}^{\infty}
\dfrac{q\tau-\Omega}{q\tau-z'}e^{-\tau^2}d\tau.
$$

Таким образом, константа $\gamma$ найдена:
$$
\gamma=-\dfrac{iy}{2}q^2\dfrac{N}{v_T^3} \dfrac{J_{12}}{J_{01}},
$$
где
$$
J_{01}=\dfrac{1}{\sqrt{\pi}}\int\limits_{-\infty}^{\infty}
\dfrac{q\tau-\Omega}{q\tau-z'}e^{-\tau^2}d\tau.
$$

Теперь формулу (2.4) можно представить в виде:
$$
j_x=\dfrac{Ne^3E_y^2q}{2k_T^2p_T^2v_T\Omega}\Big[
\Omega+\dfrac{iy}{2}-\dfrac{iy}{2}q\dfrac{J_{02}}{J_{01}}\Big]J_{12}.
\eqno{(2.5)}
$$

В выражении перед интегралом из (2.4) выделим плазменную
(ленгмюровскую) частоту
$$
\omega_p=\sqrt{\dfrac{4\pi e^2N}{m}}.
$$

Получим
$$
{j_x}^{\rm long}=\Big(\dfrac{e\Omega_p^2}{k_Tp_T}\Big)
\dfrac{k{E_y^2}}{8\pi\Omega}\Big[
\Omega+\dfrac{iy}{2}-\dfrac{iy}{2}q\dfrac{J_{02}}{J_{01}}\Big]J_{12},
\eqno{(2.5)}
$$
где
$$
\Omega_p=\dfrac{\omega_p}{k_Tv_T}=\dfrac{\hbar\omega_p}{mv_T^2}
$$
-- безразмерная плазменная частота.

Равенство (2.5) перепишем в виде
$$
j_x^{\rm long}=J(\Omega,y,q)\sigma_{l,tr}kE_y^2,
\eqno{(2.6)}
$$
где $\sigma_{l,tr}$  продольно--поперечная проводимость,
$J(\Omega,y,q)$ -- безразмерная часть тока,
$$
\sigma_{l,tr}=
\dfrac{e \Omega_p^2}{p_Tk_T}=\dfrac{e\hbar}{p_T^2}
\Big(\dfrac{\hbar \omega_p}{mv_T^2}\Big)^2=
\dfrac{e}{k_Tp_T}\Big(\dfrac{\omega_p}{k_Tv_T}\Big)^2,
$$
$$
J(\Omega,y,q)=\dfrac{1}{8\pi \Omega}
\Big[
\Omega+\dfrac{iy}{2}-\dfrac{iy}{2}q\dfrac{J_{02}}{J_{01}}\Big]J_{12}.
$$

Если ввести поперечное поле
$$
\mathbf{E}_{\rm tr}=\mathbf{E}-\dfrac{\mathbf{k(Ek)}}{k^2}=
\mathbf{E}-\dfrac{\mathbf{q(Eq)}}{q^2},\qquad
{\bf kE}_{tr}=\dfrac{\omega}{c}[{\bf E,H}],
$$
то равенство (2.6) представим в инвариантной форме
$$
\mathbf{j}^{\rm long}=J(\Omega,y,q)\sigma_{l,tr}{\bf k}{\bf E}_{tr}^2
=J(\Omega,y,q)\sigma_{l,tr}\dfrac{\omega}{c}[{\bf E,H}].
$$

{\sc Замечание 1.}
Из формулы (2.5) (или (2.6)) видно, что при $y=0$ (или
$ \nu=0$) т.е. когда частота столкновений стремится к нулю и плазма становится
бесстолкновительной ($z\to \Omega, z '\to \Omega $),
эта формула в точности переходит в соответствующую формулу из
работы  \cite{Lat8} для бесстолкновительной плазмы
$$
{j_x}^{\rm long}=\sigma_{\rm l,tr}k{E_y^2}
\dfrac{1}{8\pi \sqrt{\pi}}\int\limits_{-\infty}^{\infty}
\dfrac{e^{-\tau^2}d\tau}{(q\tau-\Omega)^3}.
$$

Перейдем к рассмотрению случая малых значений волновых чисел.
Из выражения  (2.5) при малых значениях волновых чисел получаем

$$
{j_x}^{\rm long}=-\dfrac{\sigma_{\rm l,tr}k{E_y^2}}{8\pi \Omega zz'}=
-\dfrac{\sigma_{\rm l,tr}k{E_y^2}}
{8\pi \Omega (\Omega+iy)(\Omega+\dfrac{iy}{2})}=
$$
$$
=-\dfrac{e}{8\pi m\omega}
\Big(\dfrac{\omega_p}{\omega}\Big)^2\dfrac{k{E_y^2}}
{\Big(1-i\dfrac{\nu}{\omega}\Big)\Big(1-i\dfrac{\nu}{2\omega}\Big)}.
$$

{\sc Замечание 2.} При $\nu=0$ из этой формулы в точности вытекает
соответствующая формула из \cite{Lat8} для
продольного тока в случае малых значений волновых чисел в
бесстолкновительной плазмы.

\section{Заключение}

На рис. 1 и 2 представим поведение действительной (рис. 1)
и мнимой (рис. 2) частей плотности безразмерного продольного тока
при $ \Omega=1$ в зависимости от безразмерного волнового числа
$q $ при различных значениях безразмерной частоты столкновений.
При малых и больших значениях параметра $q$ кривые {\it 1,2} и {\it 3}
сближаются и становятся неразличимыми (рис. 1). Действительная часть
имеет сначала минимум, а затем максимум. Мнимая часть плотности тока имеет один
максимум. При возрастании $q$ кривые {\it 1,2} и {\it 3}
сближаются и становятся неразличимыми (рис. 2).

На рис. 3 и 4 представим поведение действительной (рис. 3) и мнимой
(рис. 4) частей плотности продольного тока в зависимости от
безразмерных волновых чисел  $q $ в случае $ \Omega=1$
при различных значениях безразмерной частоты колебаний
электромагнитного поля в случае $y=0.01$.
При больших значениях безразмерного волнового числа
кривые {\it 1,2} и {\it 3} сближаются и становятся
неразличимыми.

На рис. 5 и 6 представим поведение действительной (рис. 5) и
мнимой (рис. 6) частей плотности продольного тока в зависимости
от безразмерной частоты колебаний электромагнитного поля $ \Omega $
в случае $y=0.01$. При возрастании безразмерного волнового числа
$q $ действительные части плотности тока сближаются и
практически совпадают (рис. 5). При малых и больших значениях
волнового числа мнимые части плотности тока сближаются и в
пределе становятся неразличимыми (рис. 6).

В настоящей работе рассмотрено влияние нелинейного характера
взаимодействия электромагнитного поля с классической
столкновительной максвелловской плазмой.

Оказалось, что наличие нелинейности электромагнитного поля
приводит к генерированию электрического тока, ортогонального к
направлению поля.

В дальнейшем авторы намерены рассмотреть задачу плазменных
колебаний и задачу о скин-эффекте с использованием квадрата
векторного потенциала в разложении функции распределения.

\clearpage

\begin{figure}[ht]\center
\includegraphics[width=16.0cm, height=9cm]{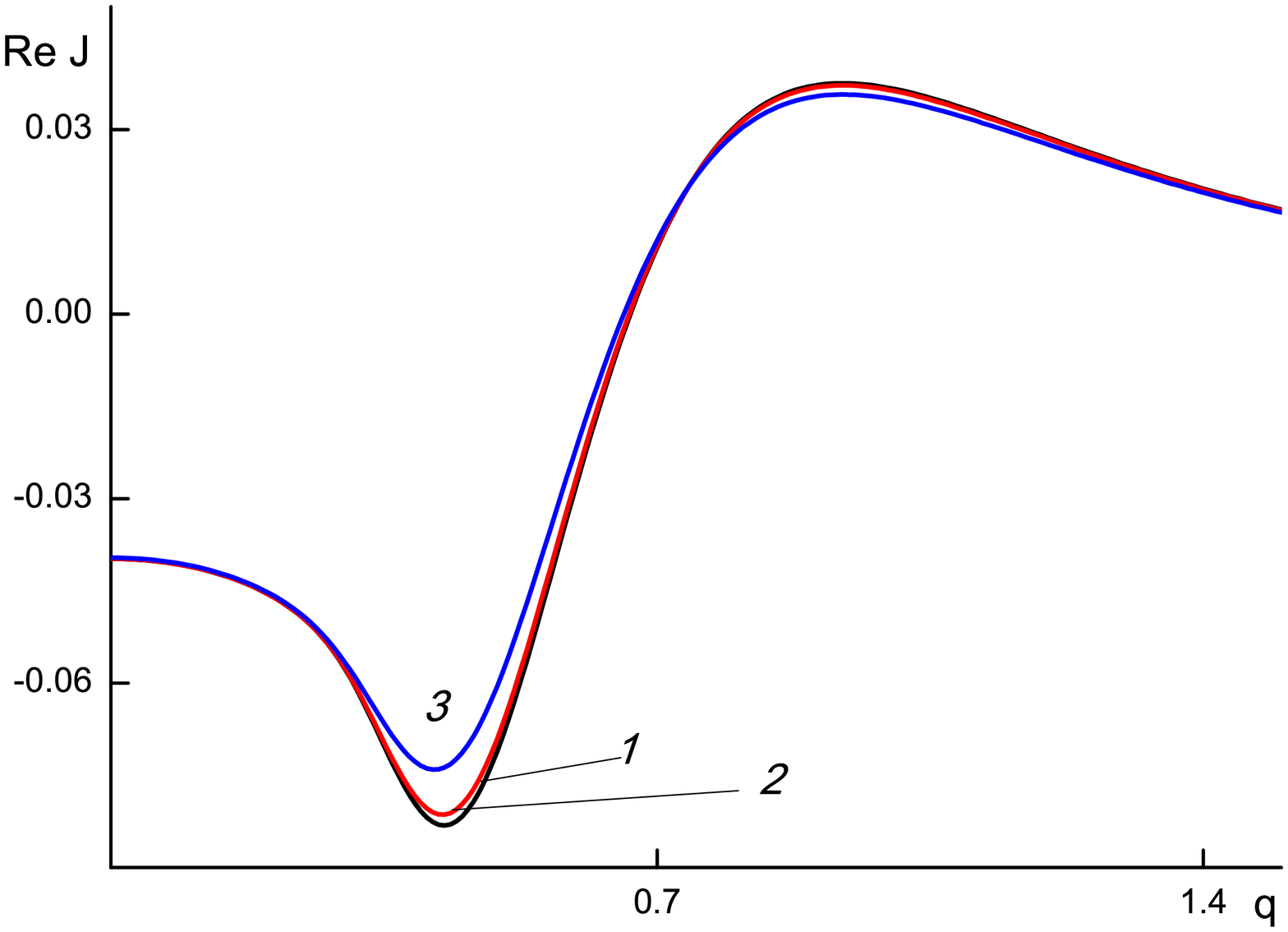}
\center{Рис. 1. Действительная часть плотности безразмерного продольного тока,
$\Omega=1$. Кривые $1,2,3$ отвечают значениям безразмерной
частоты столкновений $y=0.001, 0.01, 0.05$.}
\includegraphics[width=17.0cm, height=9cm]{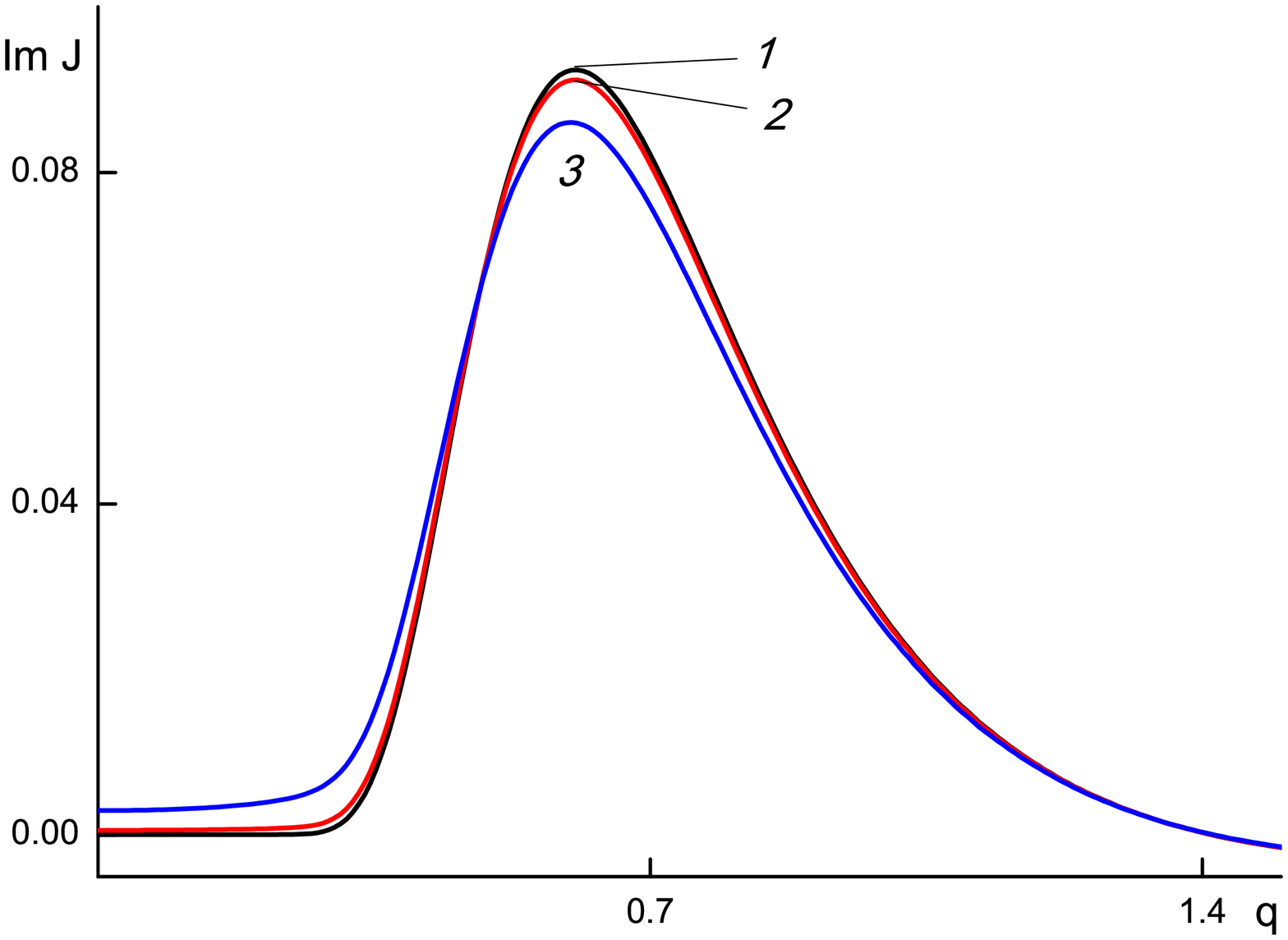}
\center{Рис. 2. Мнимая часть плотности безразмерного продольного тока,
$\Omega=1$. Кривые $1,2,3$ отвечают значениям безразмерной
частоты столкновений $y=0.001, 0.01, 0.05$.}
\end{figure}

\begin{figure}[ht]\center
\includegraphics[width=16.0cm, height=9cm]{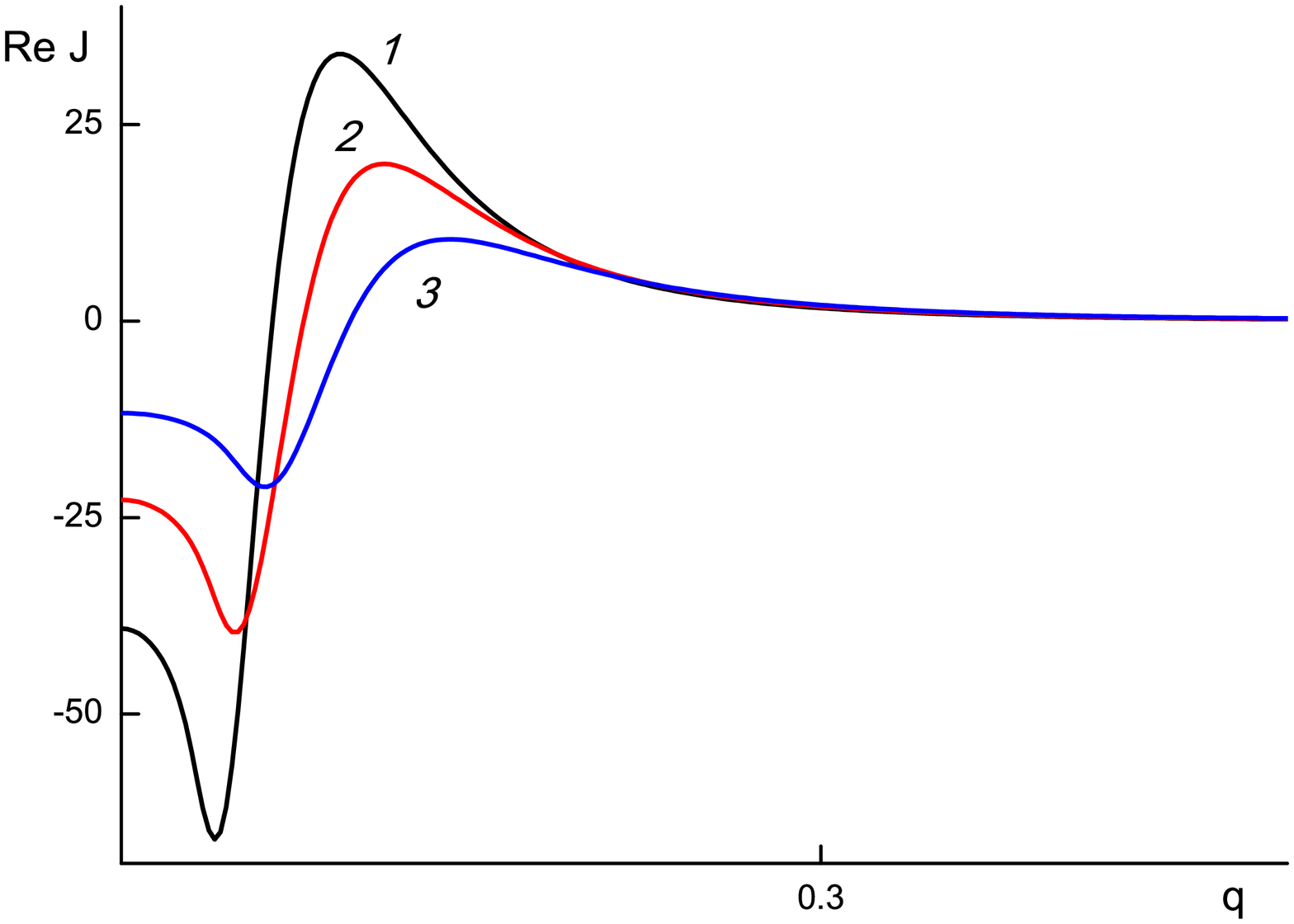}
\center{Рис. 3. Действительная часть плотности безразмерного продольного тока,
$y=0.01$. Кривые $1,2,3$ отвечают значениям
безразмерной частоты колебаний электромагнитного поля
$\Omega=0.1, 0.12, 0.15$.}
\includegraphics[width=17.0cm, height=9cm]{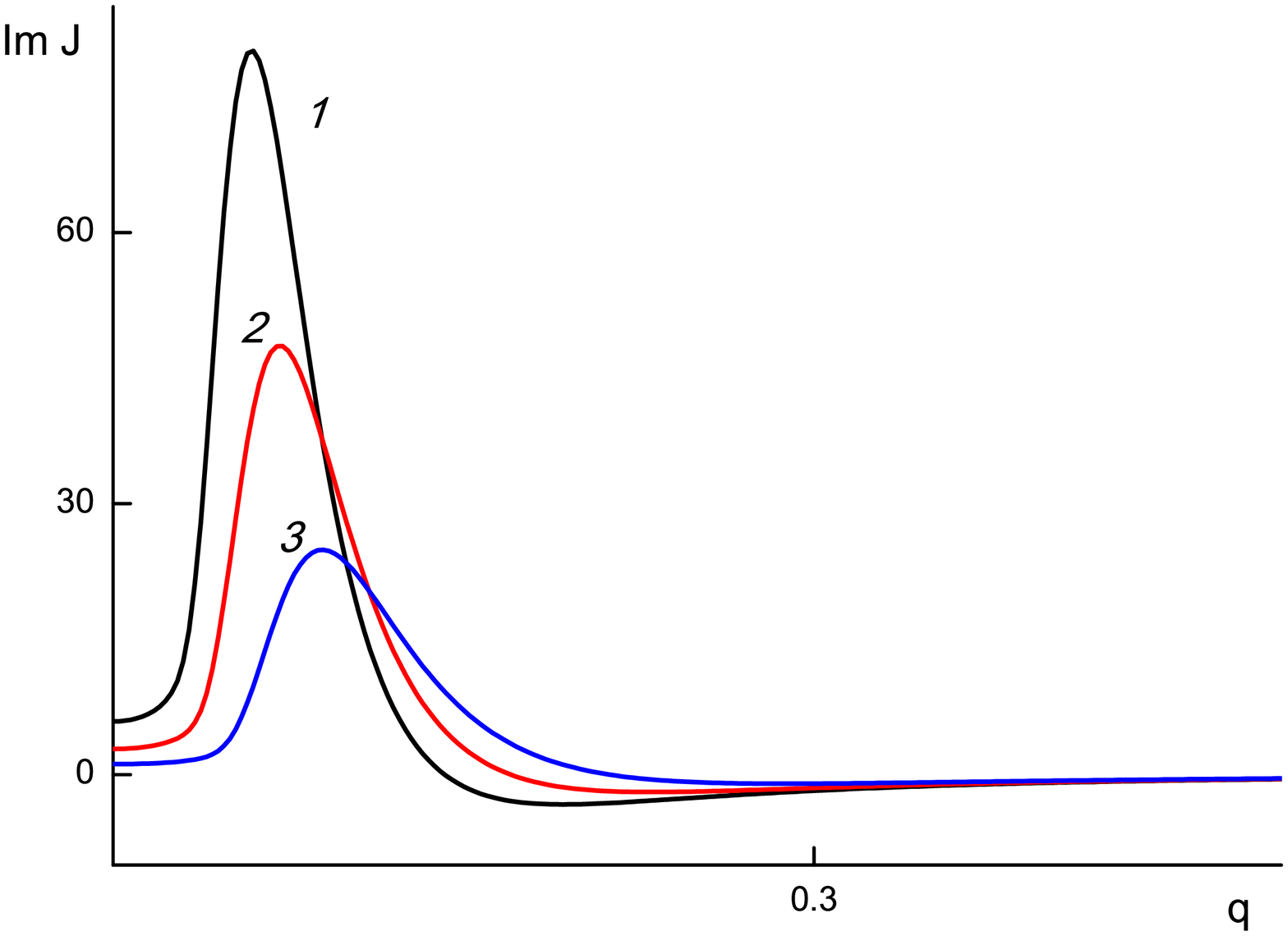}
\center{Рис. 4. Мнимая часть плотности безразмерного продольного тока,
$y=0.01$. Кривые $1,2,3$ отвечают значениям
безразмерной частоты колебаний электромагнитного поля
$\Omega=0.1, 0.12, 0.15$.}
\end{figure}

\begin{figure}[ht]\center
\includegraphics[width=16.0cm, height=9cm]{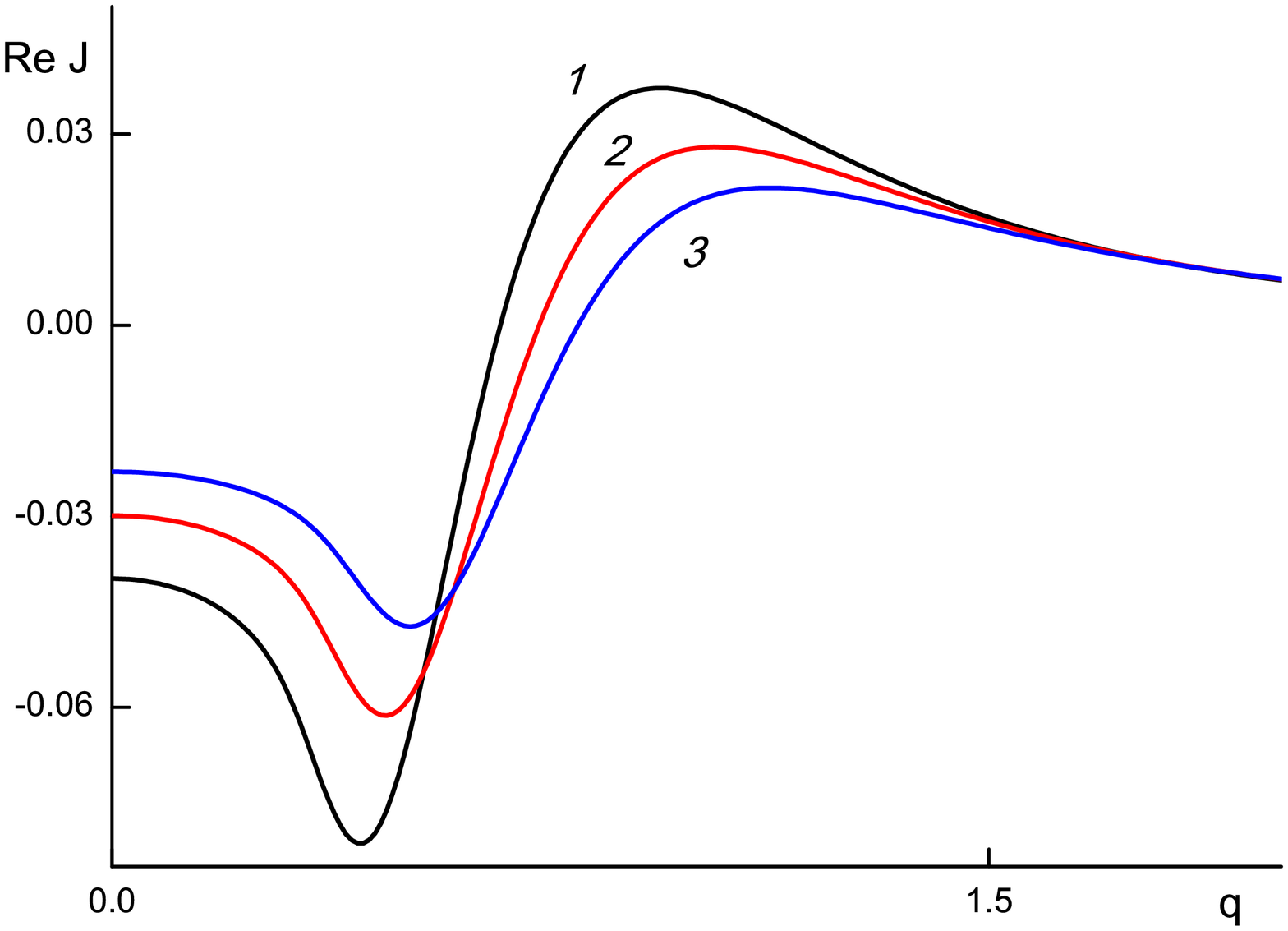}
{Рис. 5. Действительная часть плотности безразмерного продольного тока,
$y=0.01$. Кривые $1,2,3$  отвечают значениям
безразмерной частоты колебаний электромагнитного поля
$\Omega=1, 1.1, 1.2$.}
\includegraphics[width=17.0cm, height=9cm]{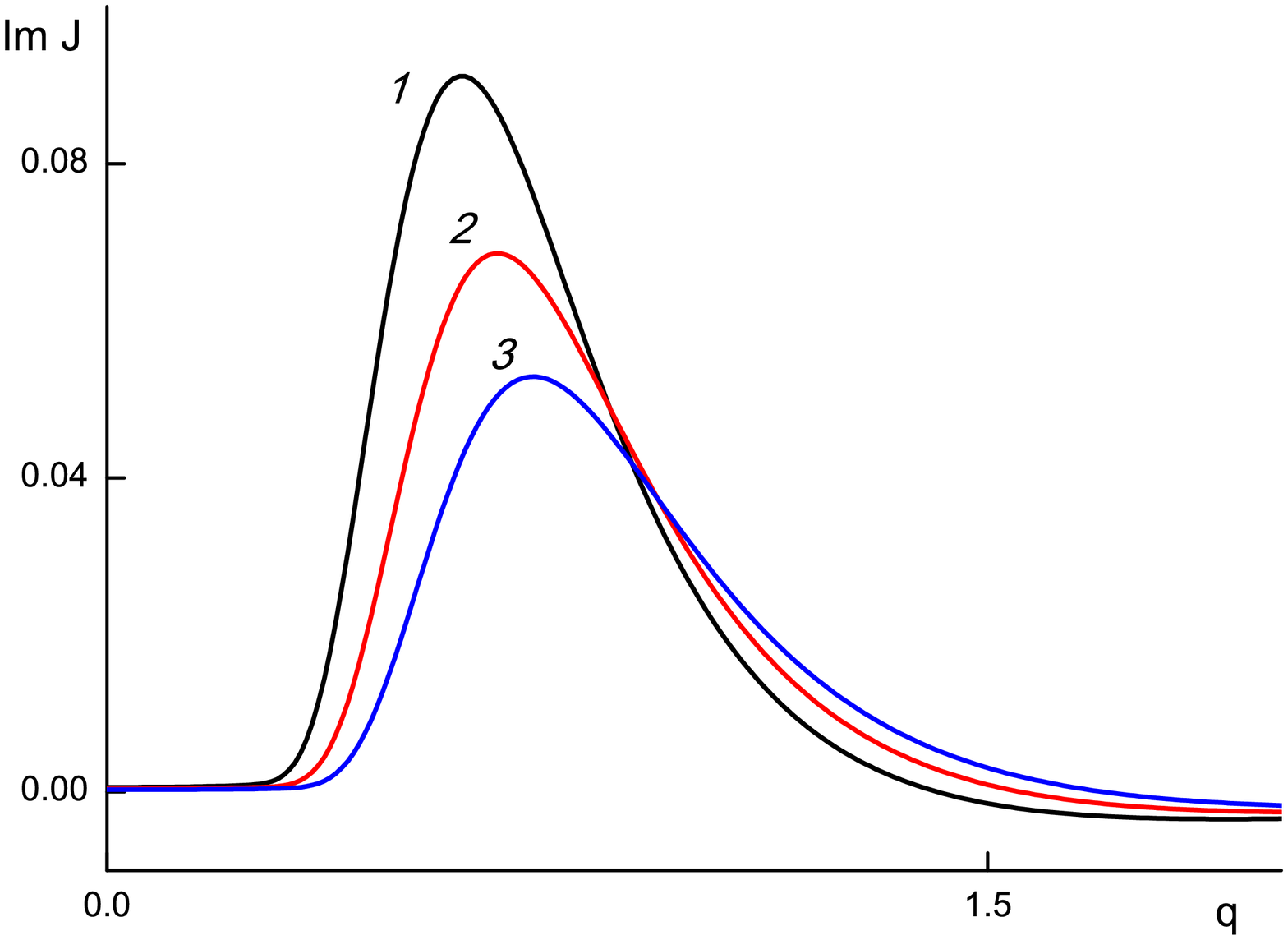}
\center{Рис. 6. Мнимая часть плотности безразмерного продольного тока,
$y=0.01$. Кривые $1,2,3$  отвечают значениям
безразмерной частоты колебаний электромагнитного поля
$\Omega=1, 1.1, 1.2$.}
\end{figure}

\end{document}